\documentclass[twocolumn,showpacs,prl]{revtex4}
\usepackage{amsmath}
\usepackage{graphicx}
\usepackage{bm}
\usepackage{color}
\usepackage{amsfonts}

\begin{document}
\title{Dynamic heterogeneity in a glass forming fluid: susceptibility, structure factor and correlation length}

\author{Elijah Flenner and Grzegorz Szamel}
\affiliation{Department of Chemistry, Colorado State University, Fort Collins, CO 80523}
\date{\today}

\begin{abstract}
We investigate the growth of dynamic heterogeneity in a glassy hard-sphere mixture 
for volume fractions up to and including the mode-coupling transition.
We use an $80\, 000$ particle system to test a new procedure to evaluate a dynamic 
correlation length $\xi(t)$: we determine the ensemble independent dynamic susceptibility $\chi_4(t)$ and use
it to facilitate evaluation of $\xi(t)$ from  the small wave vector behavior of the 
four-point structure factor. 
We analyze relations between the $\alpha$ relaxation time $\tau_{\alpha}$,
$\chi_4(\tau_{\alpha})$, and $\xi(\tau_{\alpha})$. We find that mode-coupling like power laws
provide a reasonable description of the data over a restricted range of volume fractions, 
but the power laws' exponents differ from those predicted by the inhomogeneous mode-coupling theory.
We find $\xi(\tau_\alpha) \sim \ln(\tau_\alpha)$  over the full range of volume fractions studied, which
is consistent with Adams-Gibbs-type relation.
\end{abstract}

\pacs{61.20.Lc,61.20.Ja,64.70.P-}

\maketitle

The search for a growing length scale associated with the dramatic slowing down of a supercooled liquid's dynamics 
is an active area of research. In the last decade, a growing \textit{dynamic} correlation length 
characterizing dynamic heterogeneity
was extensively studied in simulations 
\cite{Donati1999,Lacevic2003,Whitelam2004,Berthier2004,Stein2008,Flenner2009,Karmakar2009,Karmakar2010}
and experiments \cite{Berthier2005,Dalle-Ferrier2007,CrausteThibierge:2010}, and was also investigated theoretically 
\cite{Biroli2004,Biroli2006,Berthier2007p2,Berthier2007p3,Szamel2008,Szamel2010}. 
Yet there are still important and unresolved issues.  

One popular way to quantify dynamic heterogeneity is to focus on the fluctuations of particles' dynamics.
Since the dynamics is determined by two-point functions, four-point quantities 
are introduced to characterize fluctuations of dynamics, 
the so-called dynamic susceptibility $\chi_4(t)$ and associated structure factor
$S_4(q;t)$. 
Roughly speaking, $\chi_4(t)$ measures the total fluctuations of the two-point function characterizing 
particles' dynamics whereas $S_4(q;t)$ is the Fourier transform of the spatially resolved fluctuations. 
Since the total fluctuations can formally be obtained by integrating the spatially resolved ones,
one would naively expect that $\chi_4(t) = \lim_{q\rightarrow 0} S_4(q;t)$.
Alas, in a typical simulation at least some total fluctuations are not allowed (\textit{e.g.} the total number of
particles is fixed); thus $\chi_4(t)$ measured in simulations is not equal to $\lim_{q\rightarrow 0} S_4(q;t)$. 
To distinguish the susceptibility determined in an ensemble with quantity $x$ fixed we henceforth use the symbol
$\chi_4(t)|_x$. 
The difference between $\chi_4(t)|_x$ and $\lim_{q\rightarrow 0} S_4(q;t)$ makes determination of 
the dynamic correlation length $\xi(t)$ from the small wave vector behavior
of $S_4(q;t)$ more demanding. It 
necessitates \cite{Stein2008,Karmakar2010} using significantly larger systems than was customary in early simulations.

In a very interesting development, the difference between $\chi_4(t)$ and $\chi_4(t)|_x$  
was used by Berthier \textit{et al.} \cite{Berthier2005} to determine an experimentally accessible bound for $\chi_4(t)$. 
Specifically, by using the formalism developed in
Ref.~\cite{Lebowitz1967}, they showed
that $\chi_4(t) = \left. \chi_4(t) \right|_{x} + \mathcal{X}(t)$ where
$\mathcal{X}(t)$ is a correction term due to fluctuations suppressed in the constant $x$ ensemble.
Importantly, while $\left. \chi_4(t) \right|_{x}$ cannot be directly determined in experiments, the
correction term $\mathcal{X}(t)$ can. Since 
$\left. \chi_4(t) \right|_{x} > 0$, $\mathcal{X}(t)$ provides a lower bound to $\chi_4(t)$. 
Both $\left. \chi_4(t) \right|_{x}$ 
and $\mathcal{X}(t)$ were calculated using computer simulations \cite{Berthier2007p2,Brambilla2009}
and it was found that $\mathcal{X}(t)$ becomes the dominant term close to the so-called mode-coupling transition. 
However, it has not been verified that the sum of these two terms 
agrees with the independent extrapolation of $S_4(q;t)$ to $q=0$ and, somewhat surprisingly, the sum has not been
used to facilitate the evaluation of the dynamic correlation length. 

We note here two difficulties with our present understanding of dynamic heterogeneity. First,
there seems to be no consensus
regarding the scaling relation between the length measured at the $\alpha$ relaxation time, $\xi(\tau_{\alpha})$, 
and the relaxation time $\tau_\alpha$, 
even for the range of times accessible in computer simulations.
Upon approaching the mode-coupling transition almost all simulations \cite{Karmakarcom1} 
find a power law $\xi(\tau_{\alpha}) \sim \tau_{\alpha}^{1/z}$. 
However, the range of the scaling exponents reported is surprisingly large; $1/z$ varies from 0.43 \cite{Lacevic2003} 
to 0.13 \cite{SteinPHD}. 
More importantly, it is difficult to reconcile 
relationships between $\xi(\tau_{\alpha})$ and $\tau_{\alpha}$ exhibited by the simulation results 
with the experimentally determined dynamic correlation lengths \cite{Ediger2000}. Specifically, naive extrapolations
of simulational trends result in lengths that are orders of magnitude too large for relaxation times at the 
glass transition temperature \cite{extrapcom}. It has been shown \cite{Dalle-Ferrier2007,Brambilla2009} that the growth of 
the dynamic susceptibility with the relaxation time slows down near the mode-coupling transition. This 
suggests (but does not prove) a similar behavior of the correlation length.

In this Letter we address the issues mentioned in the preceding paragraphs. First, we calculate the ensemble independent
four-point susceptibility and show explicitly that it agrees very well with the extrapolation of $S_4(q;t)$ to zero wave vector. 
Next, we analyze the small wave-vector behavior of $S_4(q;t)$ and
determine $\xi(t)$. We demonstrate the practical advantage of using $\chi_4(t)$ for $\lim_{q\to 0} S_4(q;t)$. 
Finally, we analyze relations between the $\alpha$ relaxation time $\tau_{\alpha}$,
$\chi_4(\tau_{\alpha})$, and $\xi(\tau_{\alpha})$. Importantly, we find the slower-than-power law growth
$\xi(\tau_\alpha) \sim \ln(\tau_{\alpha})$.


We simulated a 50:50 binary mixture of 
hard spheres with Monte Carlo dynamics introduced 
in Ref. \cite{Brambilla2009}: the larger
sphere's diameter $d_2$ is 1.4 times larger than the smaller sphere diameter $d_1$, and the dynamics 
consists of random trial moves in a cube of length 0.1 $d_1$. We studied systems 
at fixed numbers of small and large particles ($N_1$ and $N_2$, respectively) or, equivalently,
at constant volume fraction $\phi= (N_1 d_1^3 + N_2 d_2^3) \pi/(6 V)$ ($V$ is the system's volume) 
and concentration $c=N_1/N$. We ran four trajectories with $N=80\, 000$ particles at 
$\phi = $ 0.4, 0.45, 0.5, 0.52, 0.55, 0.56, 0.57, 0.58, and 0.59, and
four trajectories with $N=10\ 000$ particles at $\phi = $ 0.54, 0.575, 0.58, 0.585, and 0.59.
We ran additional simulations to calculate derivatives with respect to $\phi$ and $c$. 
For $\phi \le 0.585$ our runs are least $100 \tau_\alpha$ long ($\tau_\alpha$ is
defined later) and for $\phi = 0.59$ we ran for $50 \tau_\alpha$ for the $80\, 000$
particle simulations and 85 $\tau_\alpha$ for the $10\, 000$ particle simulations. At each $\phi$ at least $10 \tau_\alpha$ 
were discarded for equilibration. Results are presented in reduced units: 
time in Monte Carlo steps (a Monte Carlo step
is one attempted move per particle) and lengths in the smaller particle's diameter.
This system was shown  \cite{Brambilla2009} to
reproduce very well the dynamics of an experimental glassy colloidal system.

To characterize the particles' dynamics we use the overlap function 
$w_n(t) = \theta\left(a-\left| \mathbf{r}_n(t) - \mathbf{r}_n(0) \right|\right)$
where $\theta$ is the Heaviside step function, $\mathbf{r}_n(t)$ denotes 
the position of particle $n$ at a time $t$ and $a=0.3$. The average overlap function
$F_o(t) = N^{-1} \left< \sum_n w_n(t) \right>$
encodes similar dynamic information as the self intermediate
scattering function $F_s(k;t)= N^{-1} \left< \sum_n \exp\{-i\mathbf{k}\cdot[\mathbf{r}_n(t) - \mathbf{r}_n(0)]\} \right>$. 
Here and in the following the brackets $\left< ... \right>$ denote the average over the simulational ensemble in which
the volume fraction and the composition of the system are fixed.
The sum in $F_o(t)$ 
is taken over both the large and small particles. 
We define the $\alpha$ relaxation time $\tau_\alpha$  as the time at which $F_o$
is equal to $1/e$, $F_o(\tau_\alpha) = 1/e$. We also evaluate the mean-square displacement of all the particles and the 
self-diffusion coefficient $D$. Our results for $\tau_{\alpha}$ and $D$ are consistent with those of 
Brambilla \textit{et al.} \cite{Brambilla2009}. We find that mode-coupling theory-like power laws provide good fits to the data 
for $0.55 \le \phi \le 0.58$ and the power laws' exponents are close to those predicted by the theory. 
Furthermore, we observe deviations from the power law fits for $\phi > 0.58$. 
Finally, similarly to Brambilla \textit{et al.}, we find that our results for $0.50 \le \phi \le 0.59$
can be well fitted by a function showing a stronger divergence at a higher volume fraction, $b \exp[A/(\phi_0-\phi)^2]$ with
$\phi_0 = 0.635$.

It should be emphasized that our simulations extend up to and include the
mode-coupling transition $\phi_c=0.59$ whereas previous large scale simulations \cite{Stein2008,Karmakar2010} 
covered a temperature range where mode-coupling power laws provide good fits, \textit{i.e.}\
a temperature range starting approximately 15\% above the mode-coupling temperature $T_c$. 

To characterize the heterogeneity of the system's dynamics we examine the dynamic susceptibility $\chi_4(t)|_{\phi,c}$ 
and the four-point structure factor $S_4(q;t)$,
\begin{equation}\label{eq:chi4pcdef}
\chi_4(t)|_{\phi,c} = N^{-1} \left(\left< [W(t)]^2 \right> - \left< W(t) \right>^2 \right) 
\end{equation}
\begin{equation}\label{eq:S4def}
S_4(q;t) = N^{-1}\left(\left< W(q,t) W(-q;t) \right> - \left< W(q;t) \right>^2 \right).
\end{equation}
In Eq.~(\ref{eq:chi4pcdef}) the subscript $\phi,c$ indicates that the susceptibility is calculated in the 
simulational ensemble with fixed volume fraction and concentration, and 
$W(t)$ denotes the total overlap at time $t$, $W(t) = \sum_n w_n(t)$.
In Eq.~(\ref{eq:S4def}) $W(q;t)$ is the Fourier transform of the spatially resolved overlap,
$W(q;t) = \sum_n w_n(t) \exp[-i \mathbf{q} \cdot \mathbf{r}_n(0)]$. 
We expect $\chi_4(t)|_{\phi,c} \neq \lim_{q\to 0}S_4(q;t)$ 
and we define the ensemble independent dynamic susceptibility, $\chi_4(t)$, as
\begin{equation}\label{eq:chi4def}
\chi_4(t) = \lim_{q\to 0}S_4(q;t).
\end{equation}
The difference between $\chi_4(t)$ and $\chi_4(t)|_{\phi,c}$ originates from the fluctuations of the volume fraction and the 
concentration. As recognized by Berthier \textit{et al.} \cite{Berthier2005} 
the contributions of these fluctuations to $\chi_4(t)$ can be evaluated following Ref.~\cite{Lebowitz1967}. We obtain
\begin{eqnarray}\label{eq:chi4n}
\chi_4(t) & \approx & \chi_4(t)|_{\phi,c} +  \left(\frac{\rho \pi}{6} \chi_\phi(t) \right)^2 G_1
+\frac{\rho \pi}{3} \chi_\phi(t) \chi_c(t) G_2
\nonumber \\ &&
+F_o^2(t) G_3,
\end{eqnarray}
where $\rho$ is the number density and $\chi_x(t) = \partial F_o(t)/\partial x$. 
There are other terms that contribute to Eq.~(\ref{eq:chi4n}), but they can be neglected for our system
at every studied volume fraction. 
In Eq.~(\ref{eq:chi4n}), $G_n$ are functions of the partial structure factors 
$S_{\alpha\beta}(q) = (N_{\alpha} N_{\beta})^{-1/2} 
\left< \rho_{\alpha}(\mathbf{q}) \rho_{\beta}(-\mathbf{q})\right>$ where $\rho_{\alpha}(\mathbf{q})$ is the Fourier
transform of the microscopic density of the component $\alpha$, and $\alpha,\beta=1,2$.
Explicitly, $G_1 = d_1^6 x_1 S_{11}^{(0)} +  2 d_1^3 d_2^3 \sqrt{x_1 x_2} S_{12}^{(0)} + d_2^6 x_2 S_{22}^{(0)}$
where $x_{\alpha} = N_{\alpha}/N$, and $S_{\alpha\beta}^{(0)} = \lim_{q\to 0} S_{\alpha\beta}(q)$.

\begin{figure}
\includegraphics[width=3.2in]{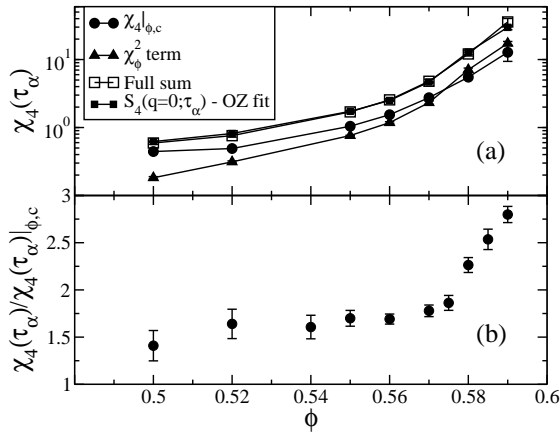}
\caption{\label{fig:sum} (a) The first two terms contributing to $\chi_4(\tau_\alpha)$ given in Eq.~(\ref{eq:chi4n}).  Also shown are
$\chi_4(\tau_\alpha)$ obtained from Ornstein-Zernicke (OZ) extrapolation of $S_4(q;\tau_\alpha)$ to $q=0$ and the sum
of all the terms in Eq.~(\ref{eq:chi4n}).
(b) The ratio $\chi_4(\tau_\alpha)/\left. \chi_4(\tau_\alpha)\right|_{\phi,c}$. 
}
\end{figure}

Shown in Fig.~\ref{fig:sum}a is the volume fraction dependence of the first two terms 
on the right hand side of Eq.~(\ref{eq:chi4n}) and the sum of all 
the terms calculated at $\tau_\alpha$. We find that the $\chi^2_\phi$ term becomes the 
dominant contribution to $\chi_4(\tau_\alpha)$ 
as $\phi$ increases \cite{Ludocom}.

To verify that $\chi_4(t)$ calculated from Eq.~(\ref{eq:chi4n}) agrees with $\lim_{q\to 0} S_4(q;t)$ and to determine $\xi(t)$ we 
need to analyze the four-point structure factor. To this end we used the 80,000 particle simulations to fit 
Eq.~(\ref{eq:S4def}) to several functions that are based on the following form,
\begin{equation}\label{fits}
S_4(q;t) = \frac{A}{1+(\xi(t) q)^2+ B^2 q^4} + \frac{C}{[1+(\xi(t) q)^2]^2}.
\end{equation}
Specifically, we used (1) the Ornstein-Zernicke (OZ) function, \textit{i.e.} we set $B=0$ and $C=0$; (2) a function suggested
by the form of a three-point susceptibility of Ref. \cite{Biroli2006}, \textit{i.e.} we set $C=0$; (3) a function suggested 
by field-theoretical considerations of Refs. \cite{Berthier2007p2,Berthier2007p3,Review2010}, 
\textit{i.e.} we set $A = \left. \chi_4(t) \right|_{\phi,c}$ and $B=0$;
(4) a function utilized in Ref. \cite{SteinPHD}, $\ln[S_4(q;t)] = \ln[A] - [\xi(t) q]^2 + D q^4$.
All procedures resulted in the same $\lim_{q\to 0}S_4(q;t)$ to within error. The results for the OZ fits are shown in 
Fig.~\ref{fig:sum} as filled squares. They agree very well with the open squares which show the right-hand-side of 
Eq.~(\ref{eq:chi4n}). 

Having verified the consistency of Eqs.\ (\ref{eq:chi4def}) and (\ref{eq:chi4n}), we now discuss the length
$\xi(\tau_{\alpha})$. Fitting procedures (1) and (2) resulted in the same length. Procedure (4) 
agreed with (1) and (2) if we restricted it to wave-vectors $q<1/\xi(\tau_\alpha)$.
As expected, procedure (3) resulted in a smaller length.  This length is approximately 1.2 times smaller than the lengths
obtained using other fitting procedures, independently of $\phi$.
Since we established the consistency of Eqs.  (\ref{eq:chi4def}) and (\ref{eq:chi4n}),
we redid the fits determining $\xi(\tau_\alpha)$ by using the right-hand-side of 
Eq.~(\ref{eq:chi4n}) for $\lim_{q\to 0}S_4(q;\tau_\alpha)$. This improved the quality 
of the fits and reduced the uncertainty in $\xi(\tau_\alpha)$. As a final revision, since the OZ function
only provided a good fit for $\xi(\tau_\alpha) q < 1.5$, we restricted the fits to values where $q < 1.5/\xi(\tau_\alpha)$. 

\begin{figure}
\includegraphics[width=3.2in,trim= 0 0 0 150,clip=true]{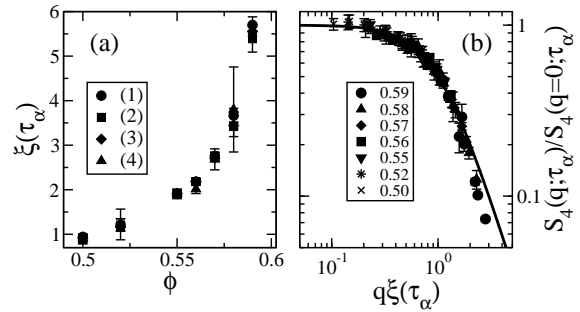}
\caption{\label{fig:sfour}(a) $\xi(\tau_\alpha)$ 
versus $\phi$ determined from the different fitting procedures discussed in the text; 
the labels denote the fitting procedure. The points
are statistically the same, and the large error bars for $\xi(\tau_\alpha)$ are due to the 
fits to $\ln[S_4(q;\tau_\alpha)]$. (b) Scaling plot of $S_4(q;\tau_\alpha)/S_4(q=0;\tau_\alpha)$ versus $q \xi(\tau_\alpha)$. 
The solid line is the Ornstein-Zernicke function used for the fits. }
\end{figure}

In Fig.~\ref{fig:sfour}a we show the results of fitting procedures (1), (2), and (4). 
We also show 1.2$\xi(\tau_\alpha)$ when $\xi$ is obtained from procedure (3). 
All these fits produce indistinguishable results. 
Finally, in Fig.~\ref{fig:sfour}b we show a scaling plot of $S_4(q;t)$ shown with the OZ function (solid line). We find
excellent overlap for every volume fraction and we observe deviations
from the OZ form only for $\xi q > 1.5$. We conclude that using the right-hand-side of 
Eq.~(\ref{eq:chi4n}) for $\lim_{q\to 0}S_4(q;\tau_\alpha)$ allows us to reliably determine $\xi(\tau_{\alpha})$.

\begin{figure}
\includegraphics[width=3.2in,trim=0 0 0 150,clip=true]{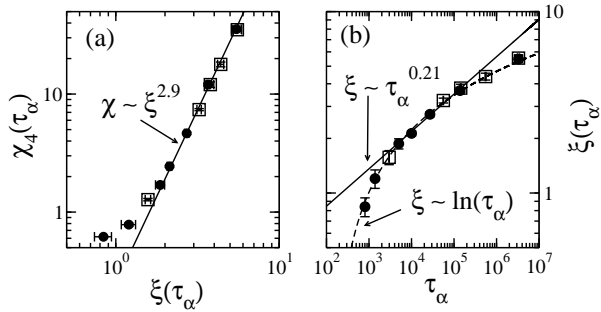}
\caption{\label{fig:chilength} (a) The dynamics susceptibility $\chi_4(\tau_\alpha)$ versus the correlation length $\xi(\tau_\alpha)$.
(b) The correlation length $\xi(\tau_\alpha)$ versus $\tau_\alpha$. The straight lines are power law fits
over the range corresponding to $0.55 \le \phi \le 0.58$. The dashed line is a fit to $\xi \sim \ln(\tau_\alpha)$ 
over the whole range of $\phi$. The circles and the squares are the results from the $80\, 000$ and
the $10\, 000$ particle simulations, respectively.} 
\end{figure}

Having tested the procedure to calculate the dynamic correlation length we now demonstrate that it can be used 
to get $\xi(\tau_{\alpha})$ using a moderately large system. We find that 
the correlation lengths determined using 80,000 and 10,000 particles are virtually identical [see Fig.~\ref{fig:chilength}] .

Recently, Karmakar \textit{et al.} \cite{Karmakar2009,Karmakar2010} advocated using finite size scaling to find the dynamic
correlation length and showed that $\xi(\tau_{\alpha})$'s obtained from this procedure are consistent with those determined
from the analysis of $S_4(q;\tau_{\alpha})$ obtained from very large scale simulations (up to $351\, 232$ particles). Finite
size scaling is attractive since it does not require large simulations. However, the version
used by Karmakar \textit{et al.} utilizes ensemble-dependent quantities, $\chi_4(\tau_{\alpha})|_{T,n,c}$ ($n$ is the total
number density) and the fourth central moment of the total overlap $W(t)$. Karmakar \textit{et al.} found that in the temperature
range investigated $\chi_4(\tau_4)/\left. \chi_4(\tau_4) \right|_{T,n,c} \approx 1.4$. We believe the temperature 
independence of this ratio might be necessary for the finite size scaling procedure to work. Figure \ref{fig:sum}b shows that 
for a range of $\phi$'s the ratio $\chi_4(\tau_4)/\left. \chi_4(\tau_4) \right|_{\phi,c}$ is approximately
constant. However, upon approaching $\phi_c$ 
it increases rapidly. Thus, Karmakar \textit{et al.}'s finite size scaling procedure 
might not work at temperatures close to and below the mode-coupling transition $T_c$.

Next, we examine scaling relations between $\tau_{\alpha}$, $\chi_4(\tau_\alpha)$ and $\xi(\tau_\alpha)$. 
As shown in Fig.~\ref{fig:chilength}a, for $\xi(\tau_\alpha)\ge 2$ we find 
$\chi_4(\tau_\alpha) \sim \xi(\tau_\alpha)^{2-\eta}$ with $2-\eta \approx 2.9$. 
Furthermore, as shown in Fig.~\ref{fig:chilength}b, 
for the same range of volume fractions where we find good power law fits 
to $\tau_\alpha$ and $D$, we observe an approximate power law 
$\xi(\tau_\alpha) \sim \tau_\alpha^{1/z}$ with $1/z \approx 0.21$.  The scaling exponents disagree with the
inhomogeneous mode-coupling predictions \cite{Biroli2006} and with Ref. \cite{Stein2008}. Interestingly, $1/z$ is consistent with
some of the earlier studies \cite{Whitelam2004,Flenner2009,Karmakar2009}. Finally, we find that 
$\xi(\tau_\alpha) \sim \ln(\tau_\alpha)$ over the whole range of volume fractions studied, thus
there is a slower-than-power law increase of the dynamic correlation length with the relaxation time.

We note that Adam-Gibbs \cite{AdamGibbs} and Random-First-Order-Transition \cite{Kirkpatrick1989} theories
postulate an exponential dependence of the relaxation time, $\tau$, on a length, $\xi$, characterizing the size
of correlated regions, $\tau \sim \exp(\xi^{\psi})$, and a relation between $\xi$ and the so-called configurational
entropy $S_c$, $\xi \sim (1/S_c)^{1/(d-\theta)}$ \cite{AGtest}. A recent study \cite{Cammarota2009} confirmed these relations
(although somewhat indirectly) and found $\psi\approx 1$, which 
is consistent with our relation between $\tau_{\alpha}$ and the dynamic correlation length.
This suggest an intriguing connection between 
the length characterizing the size
of correlated regions and the dynamic heterogeneity length \cite{Giovambattista} but we leave it
for a future study. We also acknowledge that 
another dynamic  length characterizing single-particle motion was also found to be a linear function 
of $\ln(\tau_\alpha)$ \cite{KSS}.

Finally, we examine the $\phi$ dependence of the correlation length. 
We find that, in the same range where $\tau_\alpha$ and $D$ are well described by power laws, 
$\xi(\tau_\alpha)$ is also well described by a power law with an exponent of $\gamma_\xi = 0.46 \pm 0.03$,
different from the mode-coupling exponent of 0.25. The mode-coupling-like fit breaks down
at the higher volume fractions. Since we found that $\tau_\alpha \sim e^{k \xi}$
and $\tau_\alpha \sim \exp\{A/(\phi_0 - \phi)^2\}$, we fit $\xi(\tau_\alpha) = \xi_0 + \tilde{A}/(\tilde{\phi}_0 - \phi)^2$. 
This function
provides a good fit for the whole range of $\phi$.
It results in $\tilde{\phi}_0 = 0.635 \pm 0.003$.

To summarize, through large scale computer simulations we verified that the procedure proposed by Berthier \textit{et al.}\
results in $\chi_4(t)$ that agrees very well with the independent extrapolation $\lim_{q\to 0} S_4(q;t)$. 
This allowed us to propose a new, computationally easier
procedure to evaluate the dynamic correlation length.
Importantly, we find a slower-than-power law growth of $\xi(\tau_{\alpha})$ with $\tau_{\alpha}$.


We thank L. Berthier for discussions and L. Berthier, G. Biroli, and D. Reichman for 
comments on the paper. We gratefully acknowledge the support of NSF Grant No.\ CHE 0909676.

\end{document}